\documentclass{phbauth}
\usepackage{graphicx}

\begin{document}

\begin{frontmatter}

\title{Luminescence from laser-created bubbles in cryogenic liquids}

\author{Ohan Baghdassarian},
\author{Bernd Tabbert},
\author{Gary A. Williams\thanksref{thank1}}

\address{Department of Physics and Astronomy, University of California,
Los Angeles, CA 90095, USA}

\thanks[thank1]{Corresponding author. E-mail: gaw@ucla.edu}

\begin{abstract}
A luminescence pulse has been observed from a laser-created bubble
in liquid nitrogen and liquid argon at the first collapse point of
the bubble.  An unusual feature is that the width of the
pulse is of order 100-1000 ns, much longer than the 2-8 ns pulses
observed when the same experiment is carried out with a water
sample.

\end{abstract}


\end{frontmatter}
Light emission from bubbles acoustically trapped in water has been
intensively studied over the last few years \cite{sbsl}, known as
single-bubble sonoluminescence (SBSL). Since the early 70's there
have been sketchy reports in the literature that luminescence
could also be observed from bubbles in liquid nitrogen
\cite{jltp,GOL71}. However, initial attempts to observe SBSL in
liquid nitrogen in our lab were unable to observe any light
emission from trapped helium bubbles \cite{BAG97}.  In this paper
a different technique to generate luminescing bubbles in liquid
nitrogen and liquid argon is described and compared to similar
measurements in water \cite{BAG99,OHL98}.

The bubbles are created by a Nd:YAG laser in a stainless steel
sample cell mounted inside an optical cryostat, which can be
pressurized up to 35 bar at temperatures between 65 and 90 K.
Ultra-pure argon or nitrogen (5N) is condensed into the cell. The
light emitted by an argon-ion laser passes through two sets
of quartz windows and backlights the focus region of the YAG
pulse. The attenuation of the beam by a created bubble is
monitored by a photodiode, yielding the time dependence of the
bubble radius. A photomultiplier is also used to monitor any light
emission from the bubble. For imaging purposes the photodiode is
replaced by a CCD
camera attached to a long-distance microscope, and the laser
is pulsed using an acousto-optic modulator. By these means pictures
of the bubble can be taken with exposure times as short as 500 ns.
\begin{figure}[t]
\begin{center}\leavevmode
\includegraphics[width=1.0\linewidth]{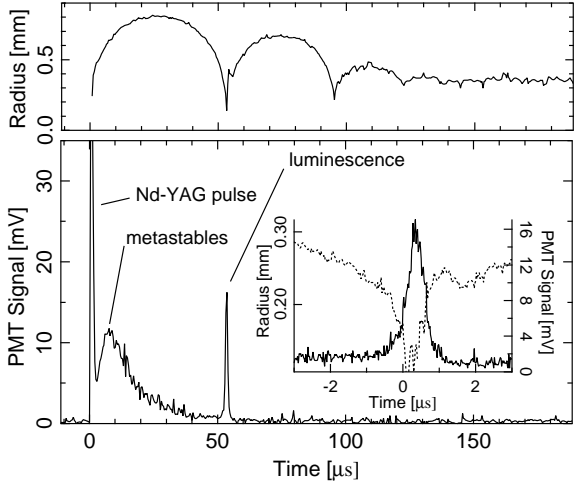}
\caption{ Dynamics and light emission of a laser-created bubble in
liquid N$_{2}$ at 66 K and 5.8 bar; upper trace: radius-time curve
showing large bounces; lower trace: PMT signal showing the plasma
emission at t = 0, the recombination of metastable species and the
luminescence peak; inset: expanded view of a similar luminescence
pulse (solid line) and radius-time curve (dashed line).
}\label{figurename1}\end{center}\end{figure}

Fig. 1 shows the dynamics and luminescence of a laser-created
bubble in liquid nitrogen.  The YAG pulse at t = 0 ionizes the
liquid at the focal point and there is a flash from the
recombining plasma.  The bubble expands to a maximum radius of
about 1 mm, while a decaying background luminescence is observed,
probably from recombining metastables such as N$_{2}$ A$^3\Sigma
_u^+$ excited molecules produced in the plasma flash. Exactly
coincident with the bubble collapse point at 52 $\mu$s, a sharp
luminescence pulse is observed on the PMT. After the
luminescence pulse the bubble rebounds with an afterbounce that is
unusually large compared to those in water \cite{sbsl,BAG99}. This
behavior appears to be similar to the very large afterbounces
observed in acoustically driven helium bubbles in liquid nitrogen
\cite{BAG97}.

The inset of Fig.  1 shows one of the luminescent pulses on an
expanded time scale; the unusual feature of these pulses compared to
those observed from laser-created bubbles in water \cite{BAG99} is
their relatively long width.  We have observed pulses with full widths
at half-maximum between 100 and 1000 ns, which appears to scale at
least roughly with the bubble size.  A similar scaling with bubble
size was observed in the water measurements, but the flash width there
was 2-8 ns, about a hundred times smaller.  The luminescence pulse in
the liquid nitrogen is consistently observed only at pressures above
3-4 bars \cite{GOL71}; at lower pressures 0.5-1 bar it is more rare,
occuring only for the largest bubbles where the pulse width is very
long (1000 ns).

Fig. 2 shows photographs of the bubble, illustrating a) an
expanding bubble pushing the thermal plume created by the focused
YAG beam before it, b) shrinking just before the first collapse, and c)
near the second collapse point, where considerable heating
of the fluid is apparent from the convection tendrils in the
region.
\begin{figure}[ht]
\begin{center}\leavevmode
\includegraphics[width=1.0\linewidth]{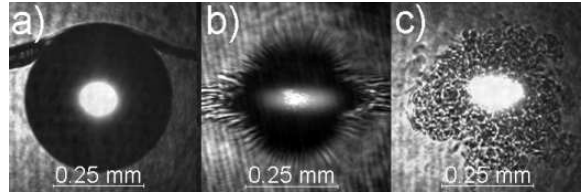}
\caption{ Photographs of bubbles created in liquid N$_{2}$ at 5.8
bar : a) bubble just after seeding, $\Delta$t = 4 $\mu$s after the
YAG pulse; b)
bubble before the first collapse, $\Delta$t = 45 $\mu$s;
c) bubble region at $\Delta$t = 90 $\mu$s, near the
second collapse. (exposure time of all pictures: 1 $\mu$s)
}\label{figurename2}\end{center}\end{figure}

The results in liquid Ar at 84 K were completely similar to those
shown here for liquid N$_{2}$.  The pulse widths were hundreds of
nanoseconds, and it was again necessary to pressurize above a few
bar to consistently see pulses at the bubble collapse point. However,
preliminary measurements in liquid He and O$_{2}$ do not show any
luminescence pulses in those liquids for pressures up to 12 bar.

This work is supported by the U. S. National Science Foundation,
DMR 97-31523. We thank M. Bernard and E. Varoquaux for initial
development of this project.  One of us (B.T.) thanks Deutsche
Akad. der Naturforscher Leopoldina for support.

\end{document}